\documentclass[a4paper,12pt]{article}
\usepackage{graphicx}
\usepackage{amsmath}
\usepackage{amssymb}
\usepackage{epsfig}

\begin{document}

\title{Numerical simulation of non-steady state neutron kinetics of the TRIGA Mark II reactor Vienna}

\author{Julia Riede, Helmuth Boeck\\{\small TU Wien, Atominstitut, A-1020 Wien, Stadionallee 2}}
{\small \date{June 14th, 2013}}
\maketitle
\begin{abstract}
This paper presents an algorithm for numerical simulations of non-steady states of the TRIGA MARK II reactor in Vienna, Austria. 

The primary focus of this work has been the development of an algorithm which provides time series of integral neutron flux after reactivity changes introduced by perturbations without the usage of  thermal-hydraulic / neutronic numerical code systems for the TRIGA reactor in Vienna, Austria. The algorithm presented takes into account both external reactivity changes as well as internal reactivity changes caused by feedback mechanisms like effects caused by temperature changes of the fuel and poisoning effects. The resulting time series have been compared to experimental results. 
\end{abstract}
\section{Introduction}
\begin{figure}[htbp]
\begin{center}
\includegraphics[width=3.5in]{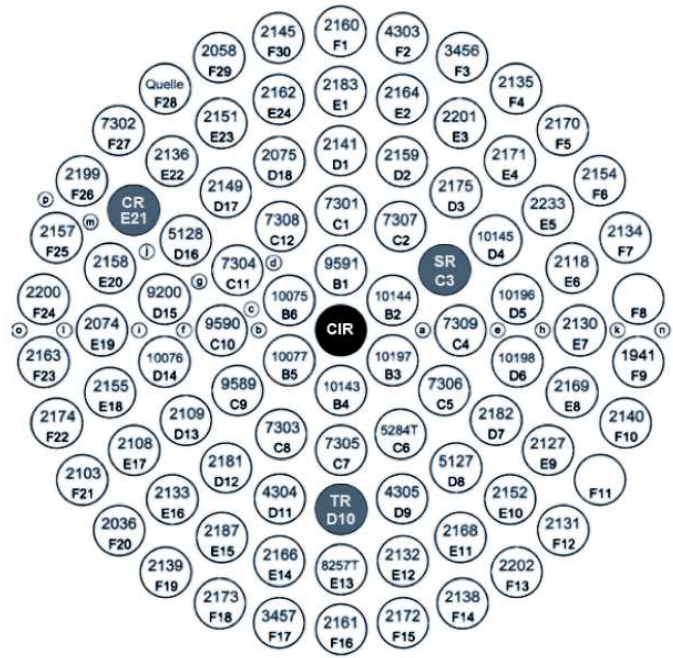}
\caption{Current core configuration with shim rod (SR), transient rod (TR), control rod (CR), startup source (\emph{Quelle}) and the central irradiation facility (CIR). The labels starting with a letter (e.g. F02) describe the position within the annular lattice; the numbers indicate the fuel type and element number.}
\label{figcore}
\end{center}
\end{figure}

The TRIGA Mark-II reactor in Vienna was built by General Atomic in the years 1959 through 1962 and went critical for the first time on March 7, 1962. Since this time the operation of the reactor has averaged at 200 days per year without any longer outages \cite{villa}\cite{trigapdf}. The reactor core currently consists of 83 fuel element arranged in an annular lattice. The current core configuration is shown in Fig. \ref{figcore}.

Publications on numerical simulation of TRIGA reactors mainly focus on steady-state and benchmarking calculations \cite{triganss1}\cite{triganss2}\cite{triganss3}\cite{triganss4}\cite{dissrustam}. If non-steady state operations are covered, only calculations using coupled thermal-hydraulic / neutronic numerical code systems like STAR or RELAP are described \cite{trigass1}\cite{trigass2}.
There has been recent work on calculation of neutron flux densities of the TRIGA reactor Vienna using MCNP \cite{dissrustam} dealing with steady-state operation focussing on the core configuration and its possible alterations. The algorithm presented in this article extends this work to enable prediction of power changes based on steady-state Monte-Carlo calculations. 

The algorithm can be used to predict changes in neutron flux (and hence reactor power) of TRIGA reactors without the usage of any coupled thermal-hydraulic / neuronic codes system. It currently only provides changes in integral neutron flux, though, and is currently not capable of calculation of local power changes.

The experimental verification is limited to integral changes in neutron flux densities as no detectors capable of measuring local neutron flux have been available at the time the experiments have been made. During the experiments, time series of integral neutron flux, reactor core temperature and inventories of Xe-135 and I-135 have been measured. 

The time series of the integral neutron flux density has been measured using a fission chamber. The temperature of the reactor core has been taken from temperature sensors built-in into three fuel elements of the reactor core displayed on the reactor control room displays. During the experiments, the time dependent inventory of Xe-135 and I-135 has been measured via gamma spectrometry, too and were found in good agreement to the values calculated by eq. \ref{eqxenon}. For the calculations below, the analytical values have been used to avoid introduction of additional error sources.
\section{Description of the algorithm}
The algorithm is based on the model presented by Emendoerfer and Hoecker \cite{emendoerfer}. Their model starts from the kinetic neutron equation for the change in neutron inventory $n(t)$ with time,
\begin{equation}
\label{eqkinetic01}
\frac{d}{dt}n(t) = \frac{\rho - \beta_{eff}}{\Lambda} n(t) + \sum\limits_{i=1}^6 \frac{\beta_{i, eff}}{\Lambda} n(0) e^{-\lambda_i t} + \int_{t'=0}^t \frac{\beta_{eff}}{\Lambda}n(t')f(t-t')dt'
\end{equation}
where
\begin{equation}
\label{eqkineticf}
f(t-t')dt' =  \sum\limits_{i=1}^6 \frac{\beta_{i, eff}}{\beta_{eff}}\lambda_i e^{-\lambda_i (t-t')}
\end{equation}
assumes $\beta_{eff,i}/\lambda$ is not time-dependent, defines the deviation from a stationary value $x(t)$ as
\begin{equation}
\label{eqdeviation}
x(t) = \frac{n(t)-n(0)}{n(0)}, \quad n(t) = n(0) \left[1+x(t)\right]
\end{equation}
with
\begin{center}
\begin{tabular}{|l|p{5cm}|l|p{5cm}|}
\hline
$\rho$ & reactivity & $\Lambda$ & mean neutron generation life time\\
\hline
$\beta_{i, eff}$ & effective fraction of delayed neutrons in group $i$& $\lambda_i$ & half-life time of precursor $i$\\
\hline
\end{tabular}
\end{center}
and presents a recursive formula for $x(t_n)$
\begin{equation}
\label{eqemnum01}
x(t_n) = \frac{\rho(t_n) + \frac{\Lambda}{\Delta t} x(t_{n-1}) + \sum\limits_{i=1}^6 F_{i,n-1} - \rho(0) }{F-\rho(t_n)}
\end{equation}
with
\begin{equation}
\label{eqemnum02}
F = \frac{\Lambda}{\Delta t} + \sum\limits_{i=1}^6 \frac{\beta_{i, eff}}{\lambda_i \Delta t} \left( 1-e^{-\lambda_i \Delta t} \right)
\end{equation}
and (for $n$ $>$ 1)
\begin{equation}
\label{eqemnum02}
F_{i,n-1} = \frac{\beta_{i, eff}}{\lambda_i \Delta t} \left( 1+e^{-2 \lambda_i \Delta t} - 2e^{-\lambda_i \Delta t}\right) x(t_{n-1}) + e^{-\lambda_i \Delta t } F_{i, n-2}
\end{equation} 
and
\begin{equation}
\label{eqemnum03}
F_{i,0} = \frac{\beta_{i, eff}}{\lambda_i \Delta t} \left[ 1-e^{-\lambda_i \Delta t} \left( 1+\lambda_i \Delta t \right) \right]
\end{equation}
Values for $\lambda_i$, $\beta_{i,eft}$ of the six delayed neutron groups are listed in table \ref{tabdelayed}.\\\\
The  method described in the equations above is applicable for continuous reactivity changes. For prompt-jump approximations at $t=0$ the very first iteration has to be calculated by \cite{emendoerfer}
\begin{equation}
x(t_0) = \frac{\rho}{\Lambda} \Delta t_0
\end{equation}
with the condition $\Delta t_0 \ll $ $\Lambda$/$\beta_{eff}$.
\section{Reactivity feedback}
The interesting part for including feedback mechanism is the calculation of the reactivity $\rho$ at a specific time $t$ (or $t_n$ if dealing with discrete values). It can be defined as
\begin{equation}
\label{rho}
\rho(t) = \rho_e(t) + \rho_i(t)
\end{equation}
with $\rho_e$ being the external reactivity (for example, moving of a control rod) and the internal reactivity $\rho_i$ (for example, changes in temperature).
\subsection{Feedback caused by changes in temperature}
The changes $z(t)$ in temperature $T$ of fuel (index $F$) and coolant (index $C$) are given by
\begin{equation}
\label{eqzb}
z_F(t) = \frac{T_F(t) - T_F(0)}{T_F(0) - T_C(0)}
\end{equation}
\begin{equation}
\label{eqzk}
z_C(t) = \frac{T_C(t) - T_C(0)}{T_F(0) - T_C(0)}
\end{equation}
Changes in temperature are calculated by using a point-model approach for fuel and coolant temperatures. Thermal conductivity from the fuel meat to the cladding is neglected and the complete heat is assumed to stay within the fuel rods. The coolant inlet temperature (index $CI$) is assumed to be constant.\\\\
Emendoerfer and Hoecker \cite{emendoerfer} show that
\begin{equation}
\label{eqzfuel}
z_F(t_n) = \frac{\Delta t \left[x(t_n) + z_C(t_n)\right] + \theta_F z_F(t_{n-1}) }{\Theta_F+\Delta t }
\end{equation}
and 
\begin{equation}
\label{eqzcoolant}
z_C(t_n) = \frac{\zeta  \Delta t z_F(t_n) +z_C(t_{n-1}) \theta }{\Delta t+\Theta_C }
\end{equation}
with $\zeta$ defined as
\begin{equation}
\zeta = \frac{T_C(0) - T_{CI}(0)}{T_F(0) - T_{CI}(0)}
\end{equation}
$\Theta_F$ and $\Theta_C$ are the fuel and coolant heat removal time constants.\\\\
As $\rho_i(t)$ is not constant but varies with internal feedback factors like temperature and poison inventory the feedback caused changes have to be calculated beforehand. This is implemented by replacing $\rho_i(t_n)$ by $\rho_i(t_{n-1})$ in an intermediate step and to correct $x(t)$ with the feedback changes calculated for this time iteration before starting the next one.
\subsection{Feedback caused by changes in reactor poison inventory}
The number of $^{135}$I nuclei $I(t)$ is given by
\begin{equation}
\label{eqi}
\frac{dI(t)}{dt} = \left(\gamma_{Sb} + \gamma_{Te}+\gamma_I\right) \Sigma_f \Phi - \sigma_I \Phi I(t)-\lambda_I I(t)
\end{equation}
with
\begin{center}
\begin{tabular}{|l|p{5cm}|l|p{5cm}|}
\hline
$\gamma_{Sb}$ & $^{135}$Sb production yield per fission & $\gamma_{Te}$ & $^{135}$Te production yield per fission\\
\hline
$\gamma_I$ & $^{135}$I production yield per fission & $\Phi$ & neutron flux\\
\hline
$\sigma_{a,I}$ & $^{135}$I absorption cross section & $\Sigma_f$ & fission cross section\\
\hline
\end{tabular}
\end{center}
Beta decay of $^{135}$I is the main source for the buildup of $^{135}$Xe. As the decay times of $^{135}$Te and $^{135}$Sb are very short compared to the decay times of $^{135}$I and $^{135}$Xe, $^{135}$Sb and $^{135}$Te can be neglected in the construction of equations \ref{eqi} and \ref{eqxe}. The number of $^{135}$Xe nuclei $Xe(t)$ is then given by
\begin{equation}
\label{eqxe}
\frac{dXe(t)}{dt} = \gamma_{Xe} \Sigma_f \Phi -\sigma_{a, Xe}\Phi Xe(t) + \lambda_I I(t) -\lambda_{Xe} Xe(t)
\end{equation}
with
\begin{center}
\begin{tabular}{|l|p{5cm}|l|p{5cm}|}
\hline
$\gamma_{Xe}$ & $^{135}$Xe production yield per fission & $\lambda_{Xe}$ &  $^{135}$Xe decay constant\\
\hline
$\sigma_{a,Xe}$ &  absorption cross section for $^{135}$Xe & & \\
\hline
\end{tabular}
\end{center}
For each time iteration in the numerical calculation the Xenon inventory has been calculated and taken into account to the internal reactivity calculation in equation (\ref{rho}). The values for the Iodine concentration were taken from the analytical solution of equation (\ref{eqi}), the values for the Xenon inventory has been calculated by solving equation (\ref{eqxe}) numerically.
\subsection{Constants and values used during the calculations}
Table \ref{tabconstants} shows the values of constants used for all subsequent calculations. 
\begin{table}[htdp]
\caption{Values of constants used for all subsequent calculations}
\begin{center}
\begin{tabular}{|l|l|l|}
\hline
Constant & Desrciption & Value\\
\hline
\hline
$\beta$ & effective delayed neutron fraction & 7.00E-03 \cite{trigapdf}\\
\hline
$l$ & mean prompt neutron lifetime [s] &  4.30E-05 \cite{trigapdf}\\
\hline
$\lambda_{I}$ & $^{135}$I decay constant [s$^{-1}$] & 2.90E-05 \cite{emendoerfer}\\
\hline
$\lambda_{Xe}$ & $^{135}$Xe decay constant [s$^{-1}$]& 2.10E-05 \cite{emendoerfer}\\
\hline
$\sigma_{a,Xe}$ & absorption cross section for $^{135}$Xe [cm$^2$]& 2.50E-19 \cite{emendoerfer}\\
\hline
$\sigma_{a,I}$ & absorption cross section for $^{135}$I [cm$^2$]&  2.20E-27 \cite{emendoerfer}\\
\hline
$\gamma_{Sb}$ & $^{135}$Sb production yield per fission& 1.50E-03 \cite{emendoerfer}\\ 
\hline
$\gamma_{Te}$ & $^{135}$Te production yield per fission & 3.13E-02 \cite{emendoerfer}\\
\hline
$\gamma_I$ &  $^{135}$I production yield per fission & 3.03E-02 \cite{emendoerfer}\\
\hline
$\gamma_{Xe}$ & $^{135}$Xe production yield per fission & 2.40E-03 \cite{emendoerfer}\\
\hline
$\Theta_F$ & Fuel feedback time constant & 3.00E-01 \cite{emendoerfer}\\
\hline
$\Theta_C$ & Coolant feedback time constant & 3.00E+00 \cite{emendoerfer} \\
\hline
\end{tabular}
\end{center}
\label{tabconstants}
\end{table}%

\begin{table}[htdp]
\caption{Delayed neutron group constants}
\begin{center}
\begin{tabular}{|c|c|c|}
\hline
Group index $i$ & $\lambda_i$ [s-1] & $\beta_i$ / $\beta_{eff}$\\
\hline
1 & 0.0124398 & 0.00021\\
\hline
2 & 0.0305082 & 0.00141\\
\hline
3 & 0.1114380 & 0.00127\\
\hline
4 & 0.3013680 & 0.00255\\
\hline
5 & 1.1363100 & 0.00074\\
\hline
6 & 3.0136800 & 0.00027\\
\hline
\end{tabular}
\end{center}
\label{tabdelayed}
\end{table}%

\section{Experimental setup}\label{exp}
Measurement of the rector power at TRIGA Vienna is done by a wide range fission fission chamber (Campbell chamber). The signal of this fission chamber has been used to record time dependent neutron flux (and hence reactor power) data. The fission chamber has been connected to an Amperemeter and the resulting signal has been recorded by a LabView application.
After 64 hours of cold shutdown state the reactor has been started up to a steady-state power level of 35W with the transient and regulating rods fully withdrawn from the core. The shim rod then has been moved from its steady state position upwards introducing the reactivity changes shown in table \ref{srtable}. 

The reactivities shown in table \label{sortable} have been taken from a previous shim rod reactivity calibration experiment (measurement of reactivity versus rod position via reactor period measurements). A plot for the resulting calibration curve can be found in section \ref{appendix}.

\begin{figure}[htbp]
\begin{center}
\includegraphics[width=4in,angle=-90]{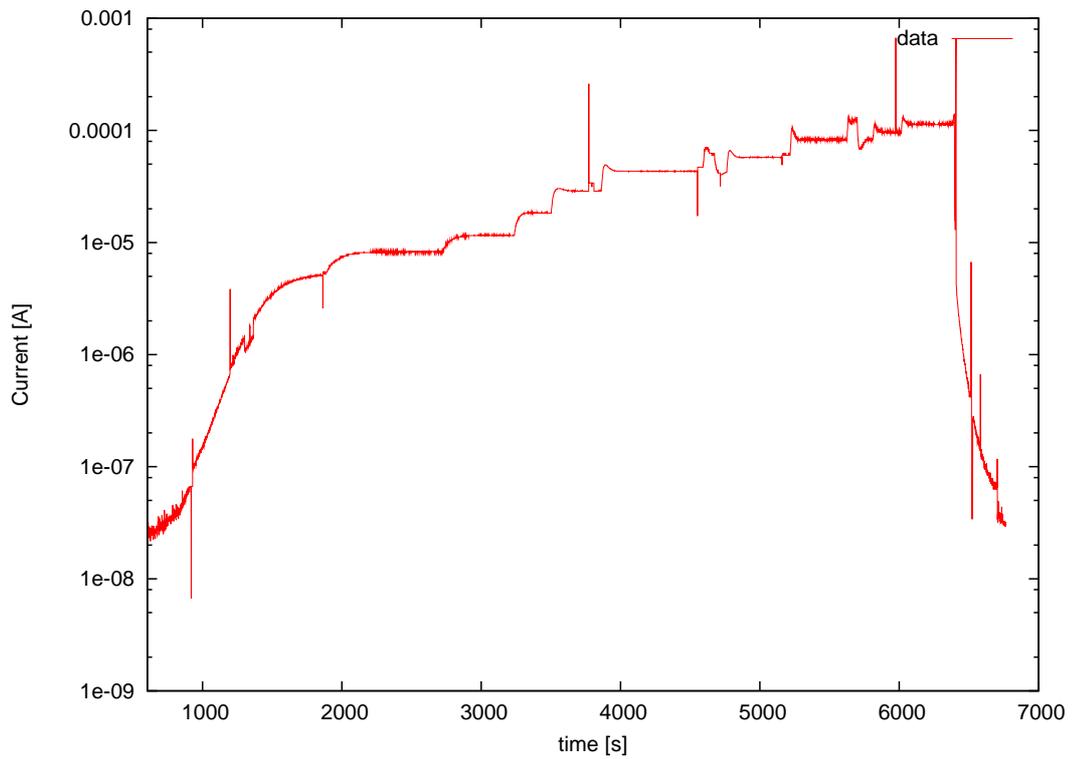}
\caption{The complete data record for the reactivity feedback measurements as an overview. Note the clearly visible temperature feedback effects showing at about an hour after measurements started. The sharp peaks are artefacts caused by changes in the Amperemeter gain settings.}
\label{figpoweroverview}
\end{center}
\end{figure}

The power levels and temperatures at each stable position have been recorded to make a fission chamber signal vs. reactor power calibration possible. An overview of the recorded, time dependent data is shown in fig. \ref{figpoweroverview}. %
\begin{table}[htdp]
\caption{Experimentally determined values for power and temperature. The reactivity has been calculated from an interpolation of the values gained by a preceding shim rod calibration experiment. The power and temperature values were taken before the reactivity has been inserted.}
\begin{center}\label{srtable}
\begin{tabular}{|c|c|c|l|c|c|c|c|}
\hline
Step & Time & Elapsed time [s] & Power &	T [$^{\circ}$C] &	 rho [c]\\
\hline
1 &	10:35:09 &	614	& 35 W $\pm$ 3\%&	35 $\pm$5	 & 0.0903\\
\hline
2 & 10:56:31 &	1896	& 9.5 kW	$\pm$ 3\%& 35 $\pm$5 &	 0.0452\\
\hline
3 &	11:10:23 &	2728 &	15.5 kW $\pm$ 3\% 	& 45 	$\pm$5  & 0.0452\\
\hline
4 &	11:18:55 &	3240 &	22 kW $\pm$ 3\%	& 50 $\pm$5 	& 0.0903\\
\hline
5 &	11:23:17 &	3502 &	50 kW$\pm$ 3\%	& 60 $\pm$5  & 0.1355 \\
\hline
6 &	11:29:17 &	3862	& 59 kW	$\pm$ 3\%& 70	$\pm$5  & 0.1807\\
\hline
7 &	11:44:16 &	4761	& 59 kW	$\pm$ 3\%& 85 $\pm$5 	 & 0.1807\\
\hline
8 &	11:51:50 &	5215	& 115 kW	$\pm$ 3\%& 100 $\pm$5   & 0.2710\\
\hline
9 &	12:01:50 &	5816	& 158 kW	$\pm$ 3\%&110 $\pm$5 	&  0.1807\\
\hline
10 &	12:05:12 &	6017	 & 189 kW	 $\pm$ 3\%& 130 $\pm$5   & 0.1807\\
\hline
\end{tabular}
\end{center}
\label{default}
\end{table}

\newpage{\pagestyle{empty}%
}
\section{Results}
\subsection{Numerical calculations}\label{num}
The method described above has been implemented in the PERL programming language and applied to various reactor states from cold shutdown status to power level conditions. Each run was started from the conditions of the stable phase of the run before including the whole new parameter set.

The algorithm has been applied to various problems defined by their fuel temperature and the external reactivities (perturbations). The interesting temperature range is between 30 C (normal shutdown conditions) and 150 C (temperature at a nominal reactor power of 250 kW). Various theoretical external reactivities were applied to states with fuel temperatures within the range mentioned above. Selected results are presented below, more can be found in section \ref{appendix}.

The effects of negative feedback, especially temperature feedback, are clearly visible: the negative feedback caused by temperature changes in the system damps the value of $x(t)$ resulting in convergence to a new steady-state position after changes in external reactivity. This is expected from inclusion of the internal reactivity according to equation \ref{rho}, otherwise the reactor would not converge to a stable power level after reactivity insertions.

Time scales of the convergence seem to be reasonable and could be verified against the experiment (see section \ref{experimentalresults}). It also shows clearly that the stronger the difference between the fuel and coolant temperatures are, the larger the amplitudes of  $x(t)$ oscillations become. This is an expected result from equations \ref{eqzfuel} and \ref{eqzcoolant}.

\begin{figure}[htbp]
\begin{center}
\includegraphics[width=3.5in,angle=-90]{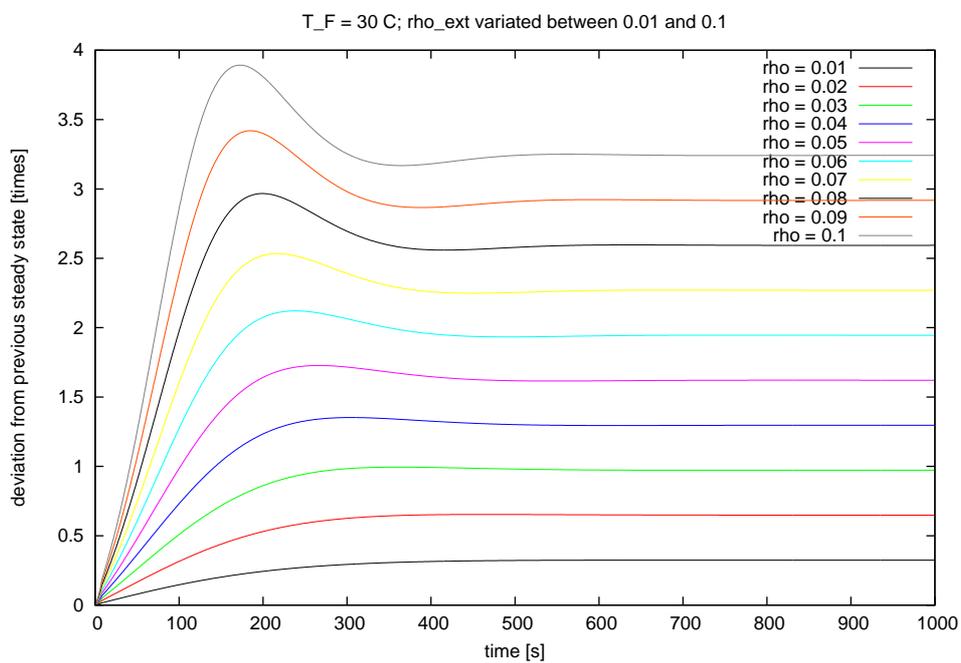}
\caption{Fuel rod temperature: 30 $^{\circ}$C, positive reactivity changes}
\label{default}
\end{center}
\end{figure}
\begin{figure}[htbp]
\begin{center}
\includegraphics[width=3.5in,angle=-90]{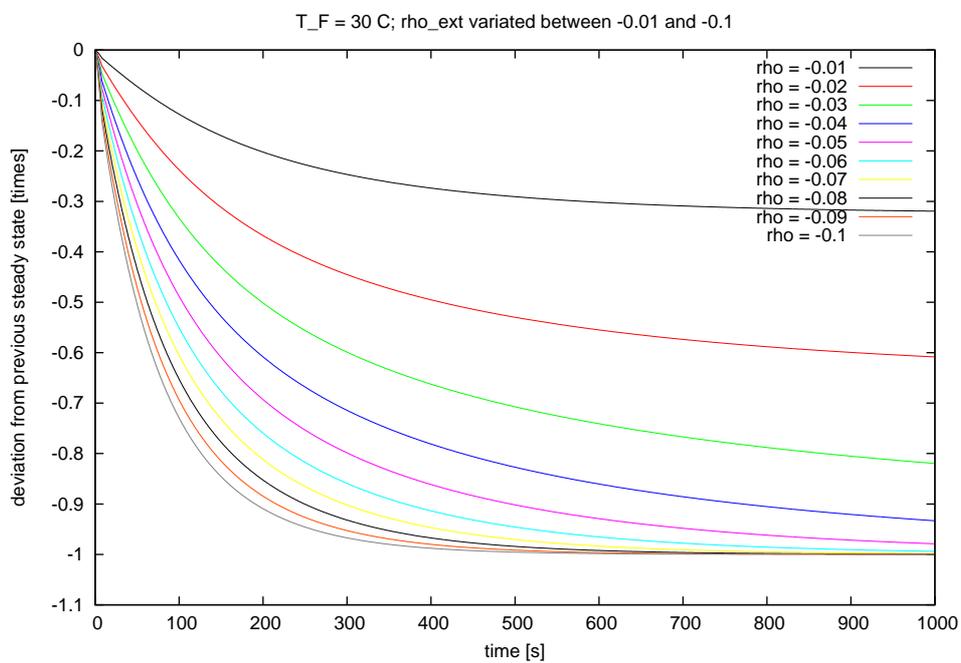}
\caption{Fuel rod temperature: 30 $^{\circ}$C, negative reactivity changes}
\label{default}
\end{center}
\end{figure}
\newpage%
\subsection{Comparison to experimental data}\label{experimentalresults}
To enable comparison of numerical and experimental data the latter has been processed to provide values for deviation from steady state values instead of absolute power levels as follows: 
Each timeframe from the previous to the current power level following a perturbation is considered as one phase. The starting steady-state power level of each phase is taken as $x(0)$ for the subsequent calculations of $x(t)$, the same procedure has been used to calculate the temperatures $T_{F,i}$ for subsequent phases. 
A new steady-state level is considered to be reached as soon as the absolute value for the calculated reactivity falls below a certain threshold. During the calculations, a threshold of $|\rho|$ $\leq$ 1.0E-04 has been used.

\begin{figure}[htbp]
\begin{center}
\includegraphics[width=3.5in,angle=-90]{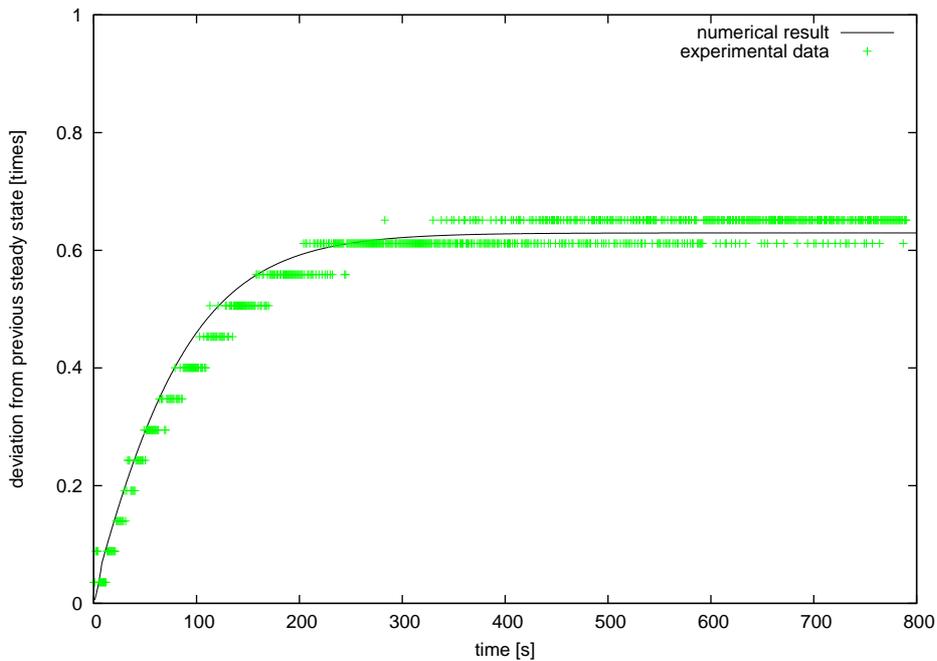}
\caption{Deviation from steady state: Experimental versus numerical results for the first step}
\label{fig-comp-30}
\end{center}
\end{figure}

The results of the numerical calculations are shown together with the experimental data in figures \ref{fig-comp-30} to \ref{fig-comp-100}.
  \newpage{\pagestyle{empty}
}\newpage
\section{Discussion}
This work has shown that a relatively simple model for reactor kinetics can be adapted to enable prediction of changes in power levels of a TRIGA reactor resulting from changes in reactivity. 

The numerical results differ from the experimental results less than 4\% for fuel temperatures up to about 55 $^{\circ}$C. As soon as the fuel temperature increases above this level, the effects of the strong negative temperature feedback coefficient of the TRIGA reactor starts to take effect which causes significant overshoots. Those are clearly visible in the numerical results, too but the shape starts to show systematic deviations to lower values during the overshoot but resulting in steady-state values in good agreement with the calculations. This is probably caused by imperfect feedback and feedback time constants ($\alpha_{TF}$, $\Theta_F$ and $\Theta_C$). The values for this constants used during the calculations are not considered to be perfect matches and need reconsideration.

Further studies are needed to include feedback caused related to the void coefficient of reactivity and reactivity changes caused by Doppler broadening. This has not taken into account as the temperature feedback coefficient clearly is the dominant factor when looking at the TRIGA reactor in Vienna.\newpage
\section{Appendix}\label{appendix}
\begin{figure}[htbp]
\begin{center}
\includegraphics[width=4in,angle=-90]{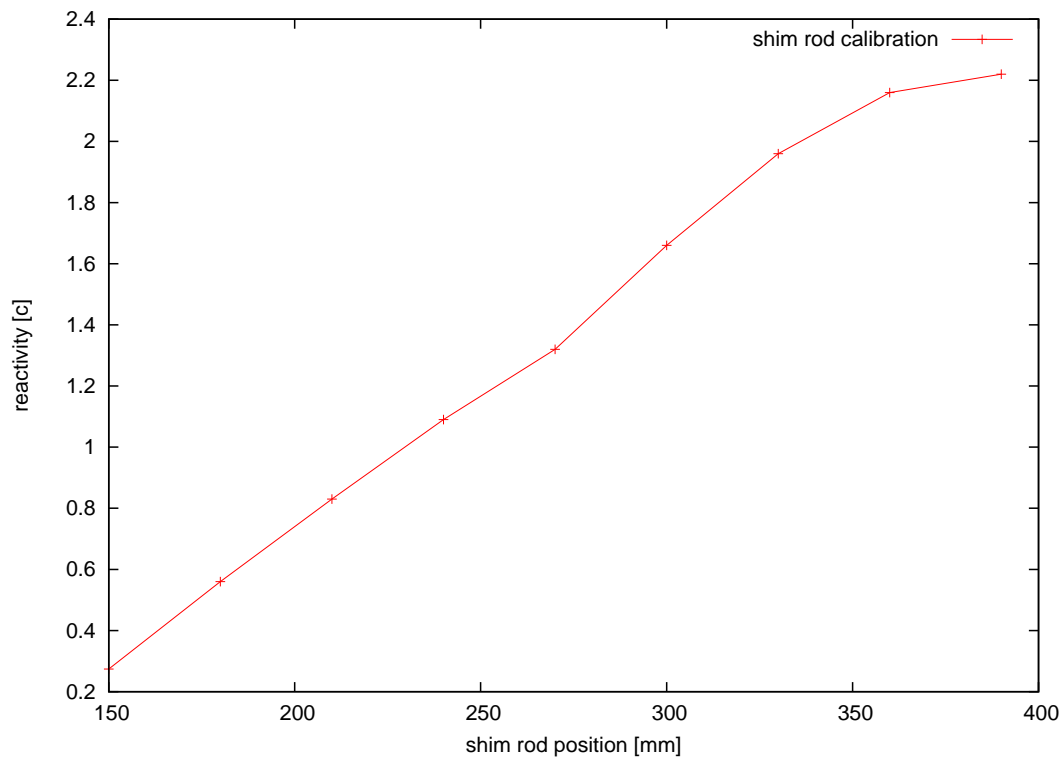}
\caption{The shim rod reactivity calibration curve obtained by experiment the day before the actual experiment has been performed. The rod is fully inserted into the core at a position of $z$ = 0 mm, positive values for rod position mean moving the rod upwards the $z$ axis.}
\label{fig-comp-100}
\end{center}
\end{figure}

\begin{figure}[htbp]
\begin{center}
\includegraphics[width=3.5in,angle=-90]{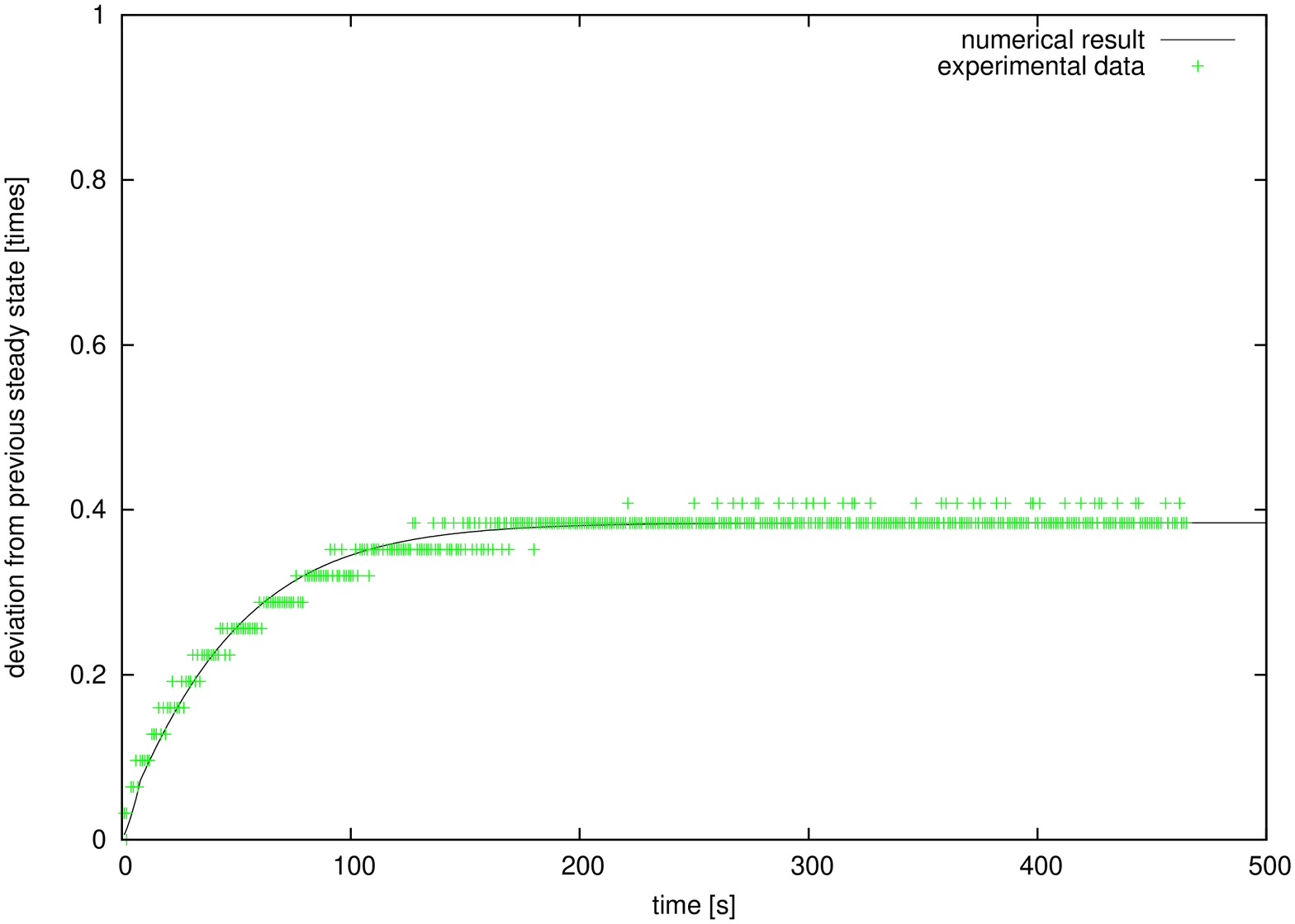}
\caption{Deviation from steady state: Experimental versus numerical results for the second step}
\label{fig-comp-40}
\end{center}
\end{figure}
\begin{figure}[htbp]
\begin{center}
\includegraphics[width=3.5in,angle=-90]{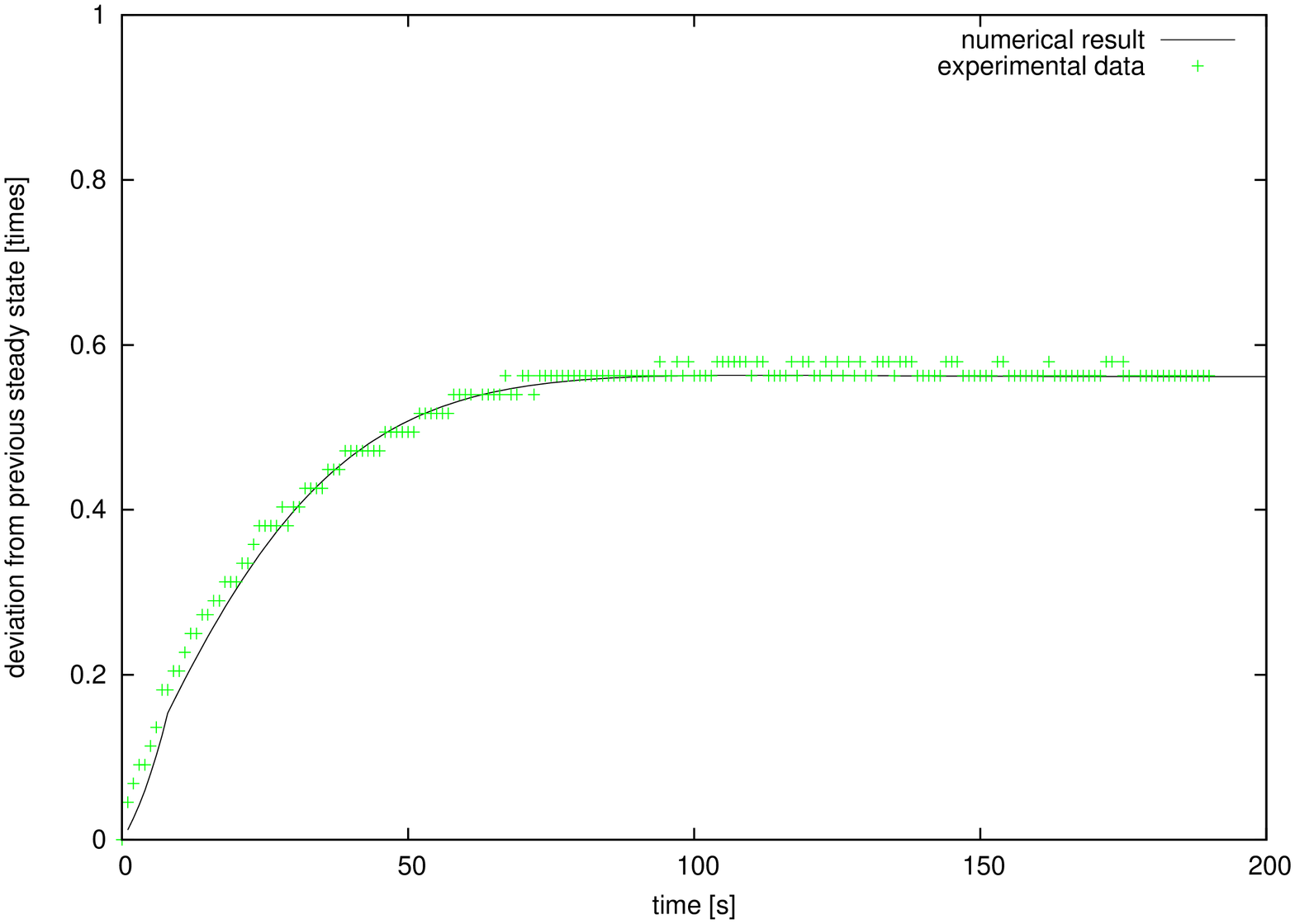}
\caption{Deviation from steady state: Experimental versus numerical results for the third step}
\label{fig-comp-50}
\end{center}
\end{figure}
\begin{figure}[htbp]
\begin{center}
\includegraphics[width=3.5in,angle=-90]{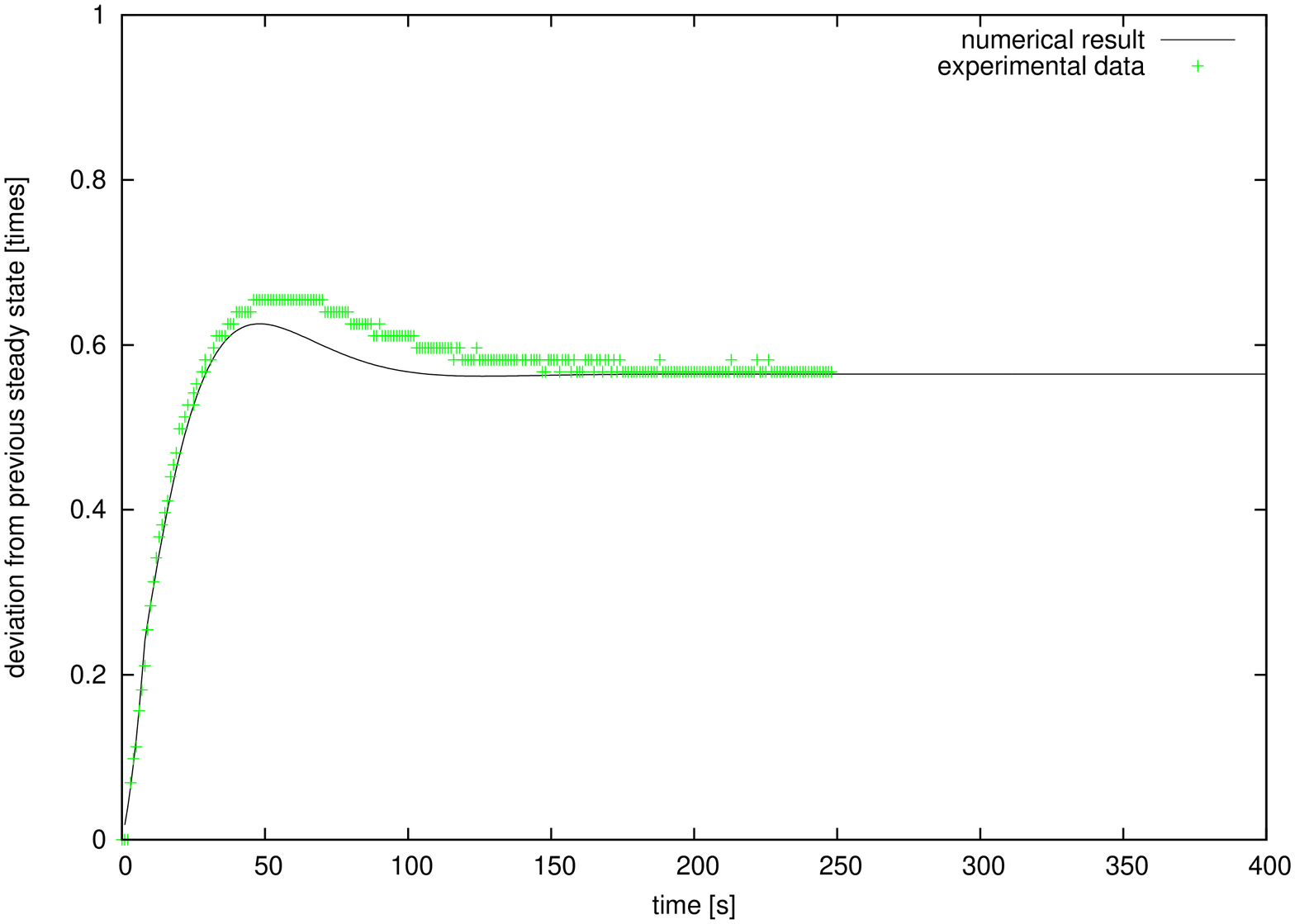}
\caption{Deviation from steady state: Experimental versus numerical results for the fourth step}
\label{fig-comp-60}
\end{center}
\end{figure}
\begin{figure}[htbp]
\begin{center}
\includegraphics[width=3.5in,angle=-90]{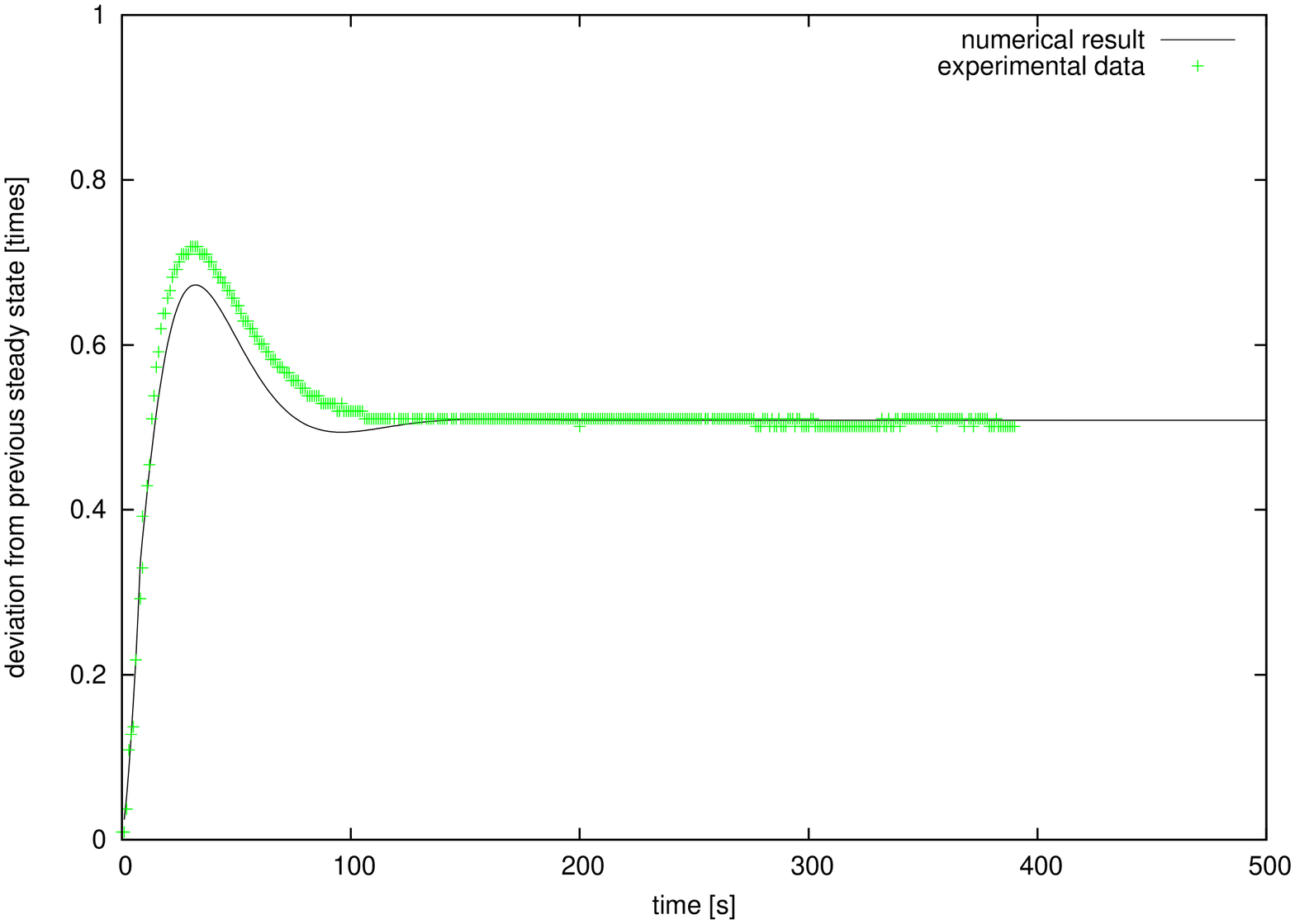}
\caption{Deviation from steady state: Experimental versus numerical results for the fifth step}
\label{fig-comp-70}
\end{center}
\end{figure}
\begin{figure}[htbp]
\begin{center}
\includegraphics[width=3.5in,angle=-90]{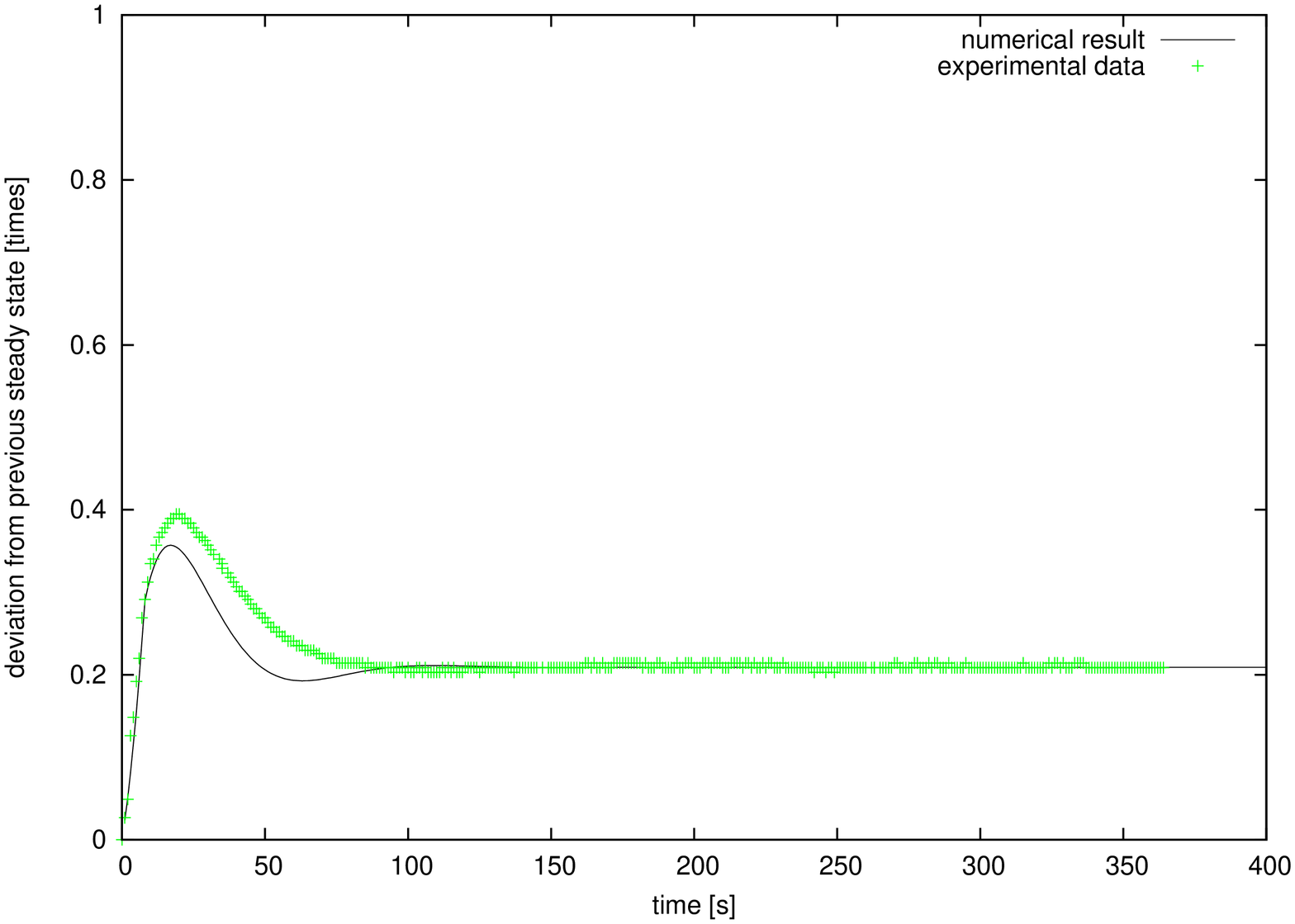}
\caption{Deviation from steady state: Experimental versus numerical results for the sixth step}
\label{fig-comp-85}
\end{center}
\end{figure}
\begin{figure}[htbp]
\begin{center}
\includegraphics[width=3.5in,angle=-90]{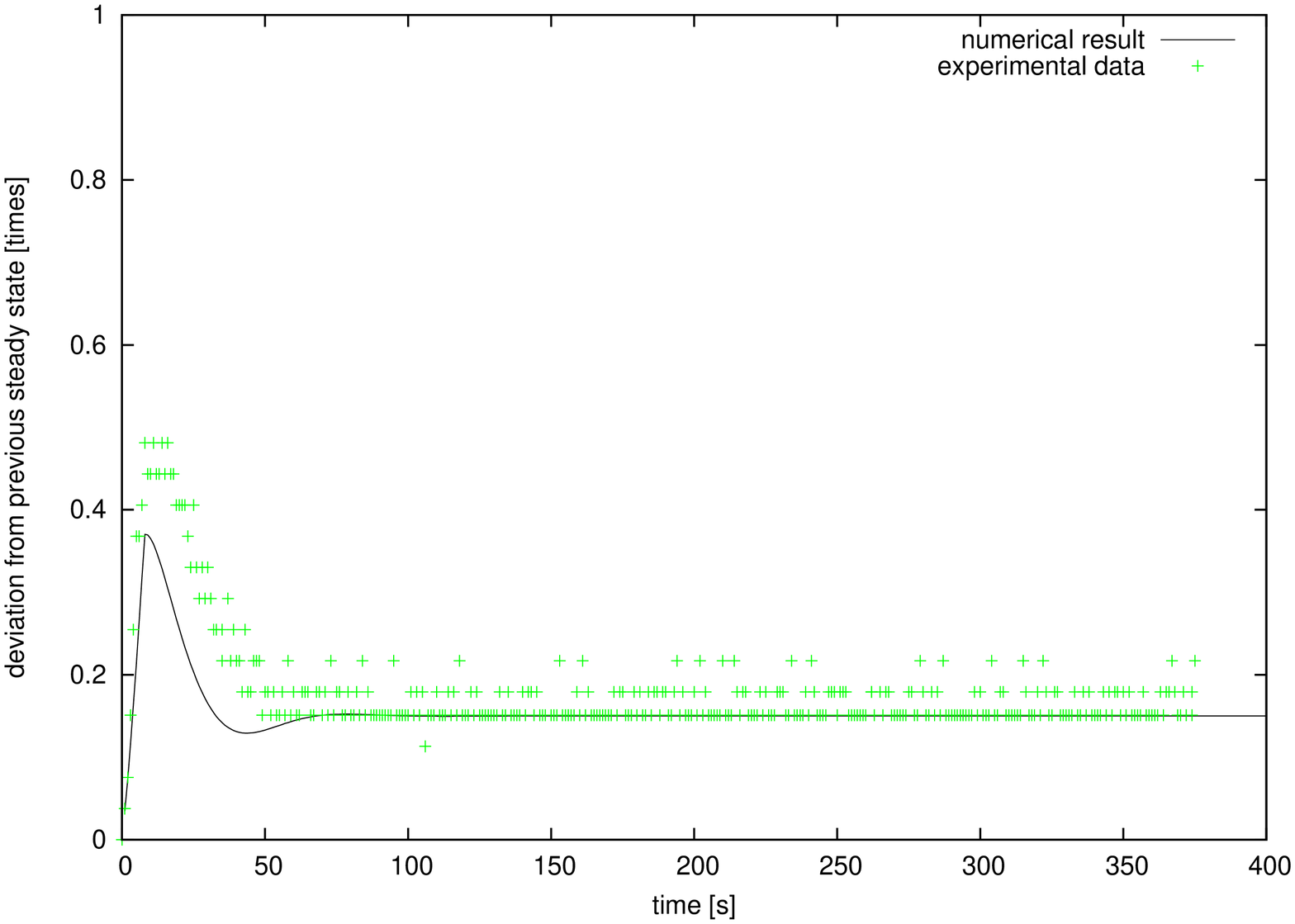}
\caption{Deviation from steady state: Experimental versus numerical results for the seventh step}
\label{fig-comp-100}
\end{center}
\end{figure}

\begin{figure}[htbp]
\begin{center}
\includegraphics[width=3.5in,angle=-90]{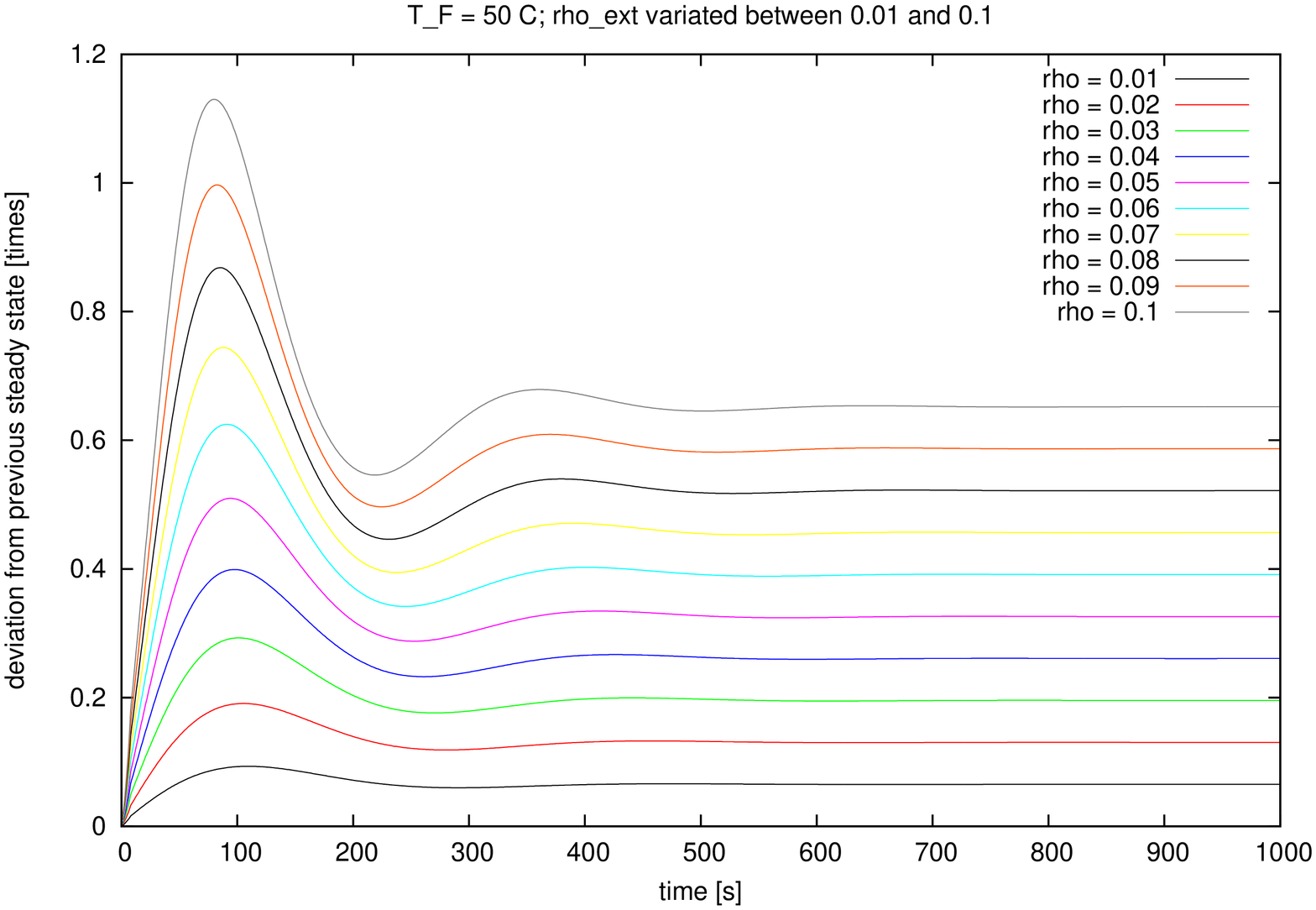}
\caption{Fuel rod temperature: 50 $^{\circ}$C, positive reactivity changes}
\label{default}
\end{center}
\end{figure}
\begin{figure}[htbp]
\begin{center}
\includegraphics[width=3.5in,angle=-90]{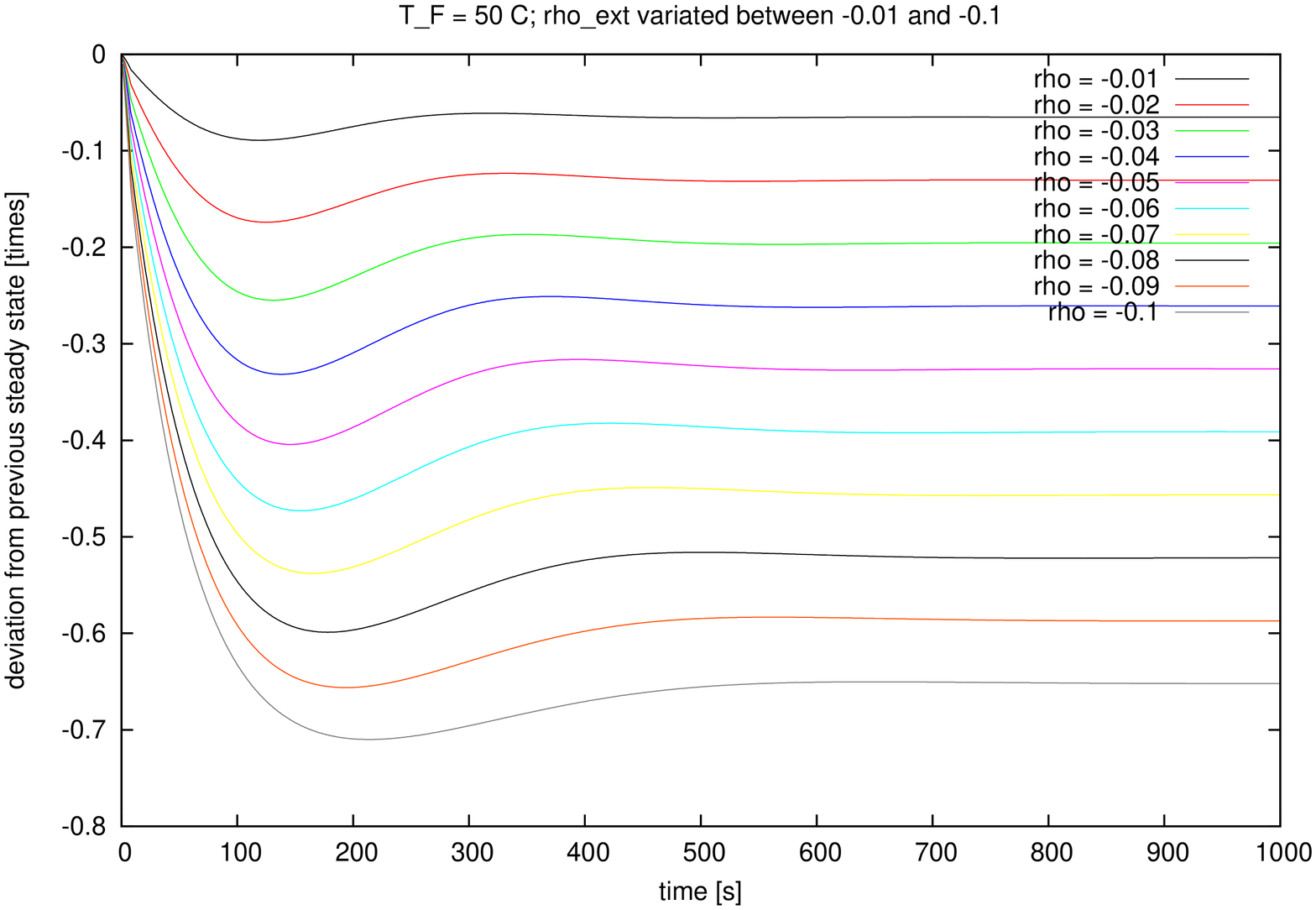}
\caption{Fuel rod temperature: 50 $^{\circ}$C, negative reactivity changes}
\label{default}
\end{center}
\end{figure}

\begin{figure}[htbp]
\begin{center}
\includegraphics[width=3.5in,angle=-90]{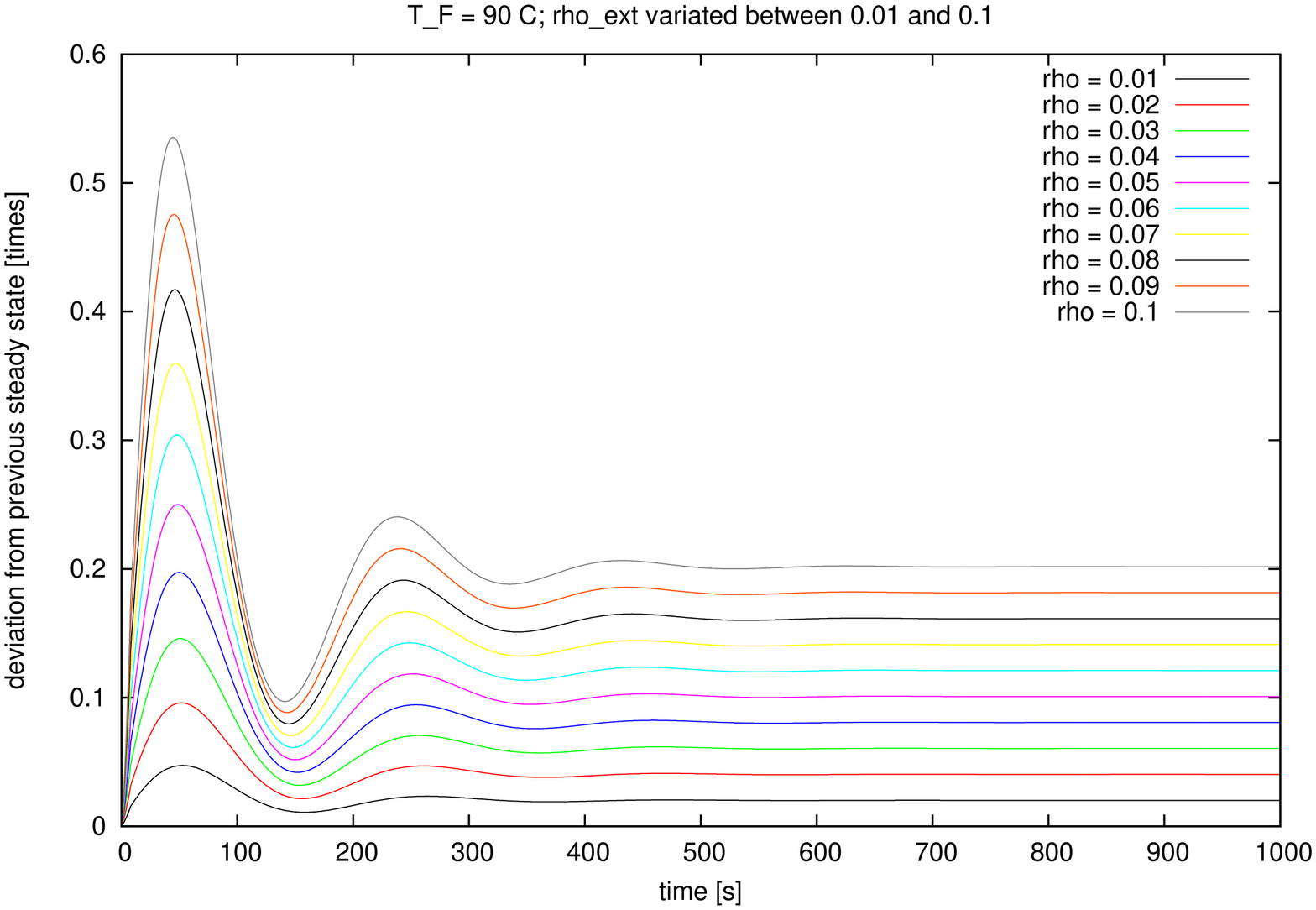}
\caption{Fuel rod temperature: 90 $^{\circ}$C, positive reactivity changes}
\label{default}
\end{center}
\end{figure}
\begin{figure}[htbp]
\begin{center}
\includegraphics[width=3.5in,angle=-90]{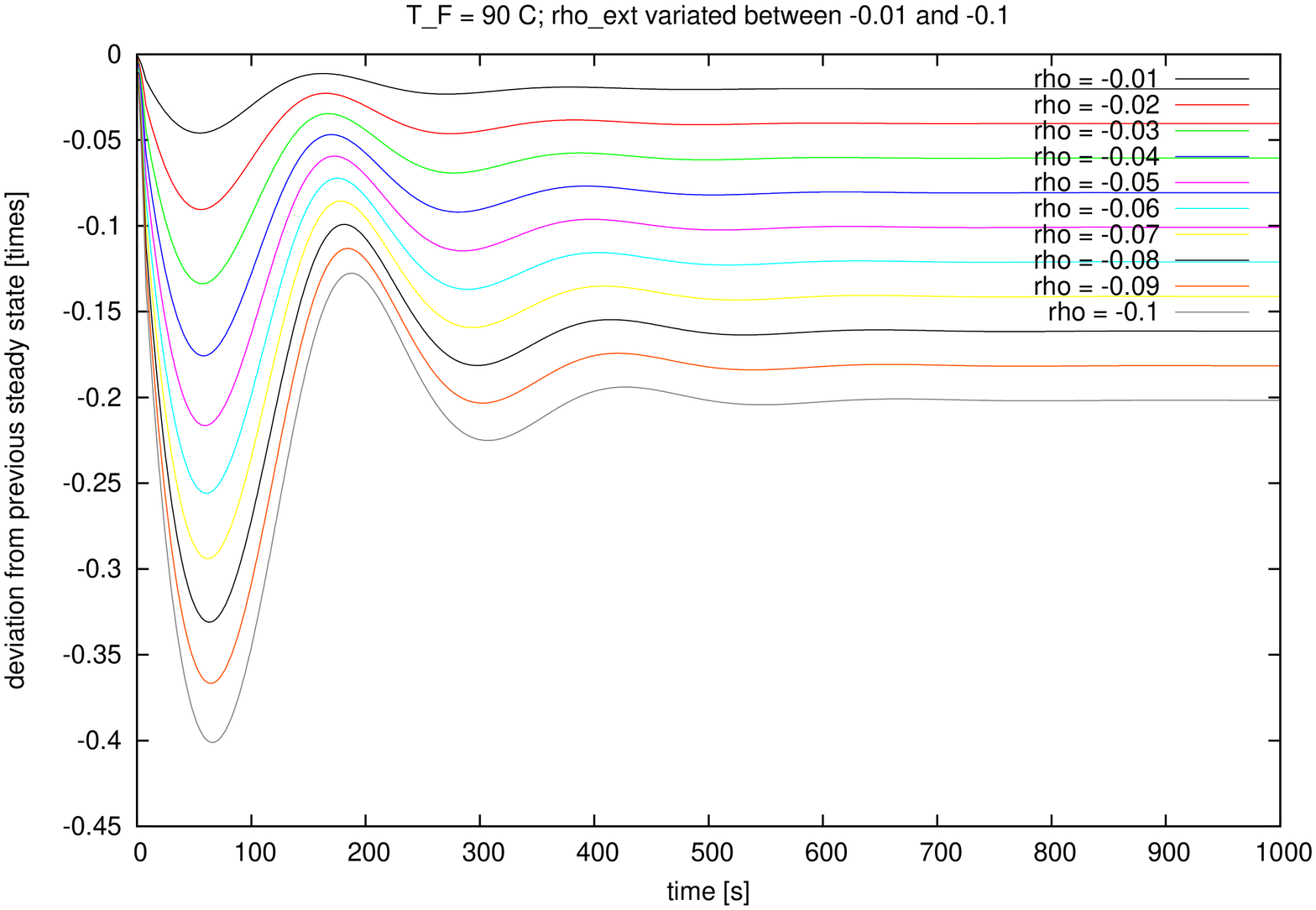}
\caption{Fuel rod temperature: 90 $^{\circ}$C, negative reactivity changes}
\label{default}
\end{center}
\end{figure}

\begin{figure}[htbp]
\begin{center}
\includegraphics[width=3.5in,angle=-90]{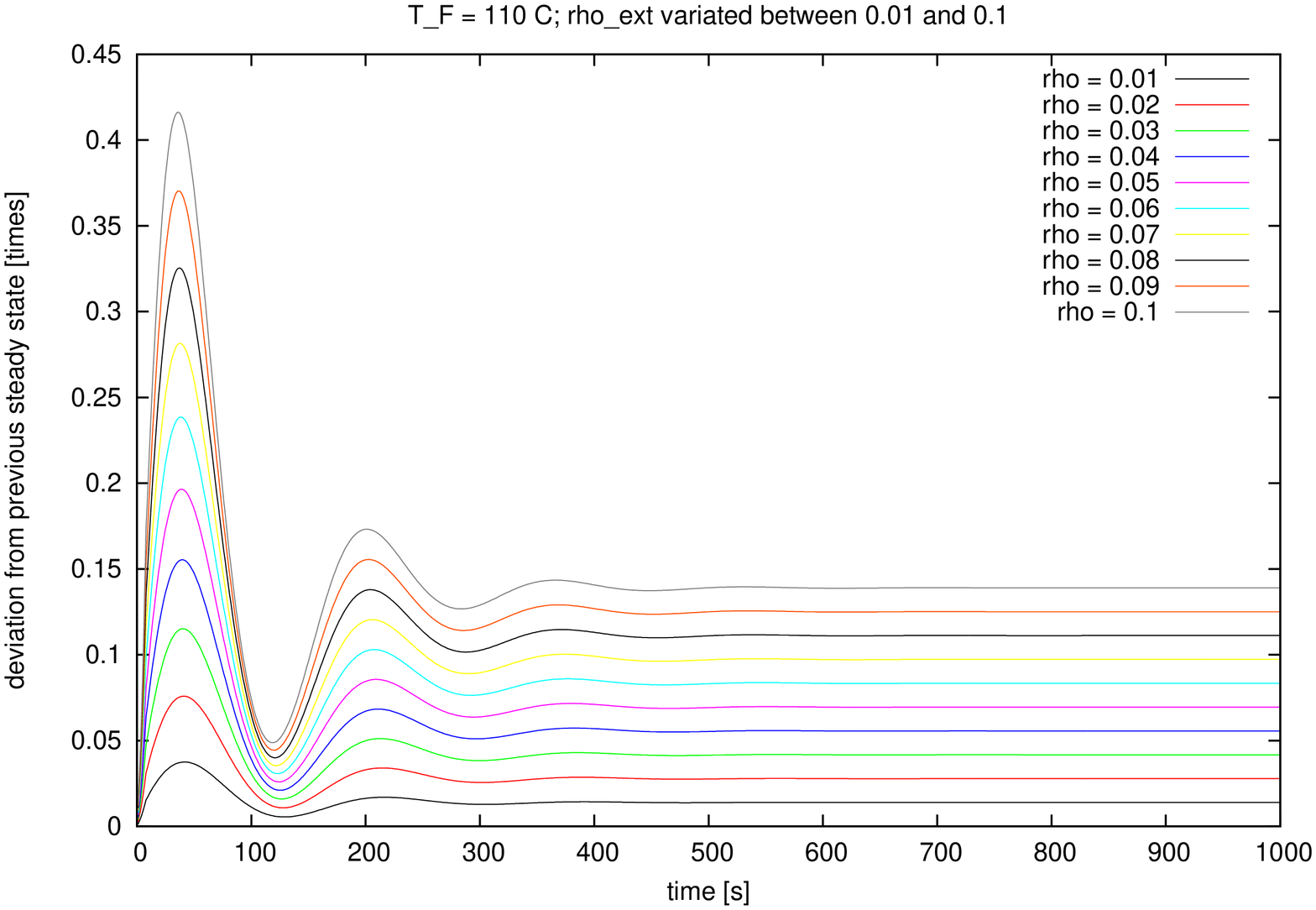}
\caption{Fuel rod temperature: 110 $^{\circ}$C, positive reactivity changes}
\label{default}
\end{center}
\end{figure}
\begin{figure}[htbp]
\begin{center}
\includegraphics[width=3.5in,angle=-90]{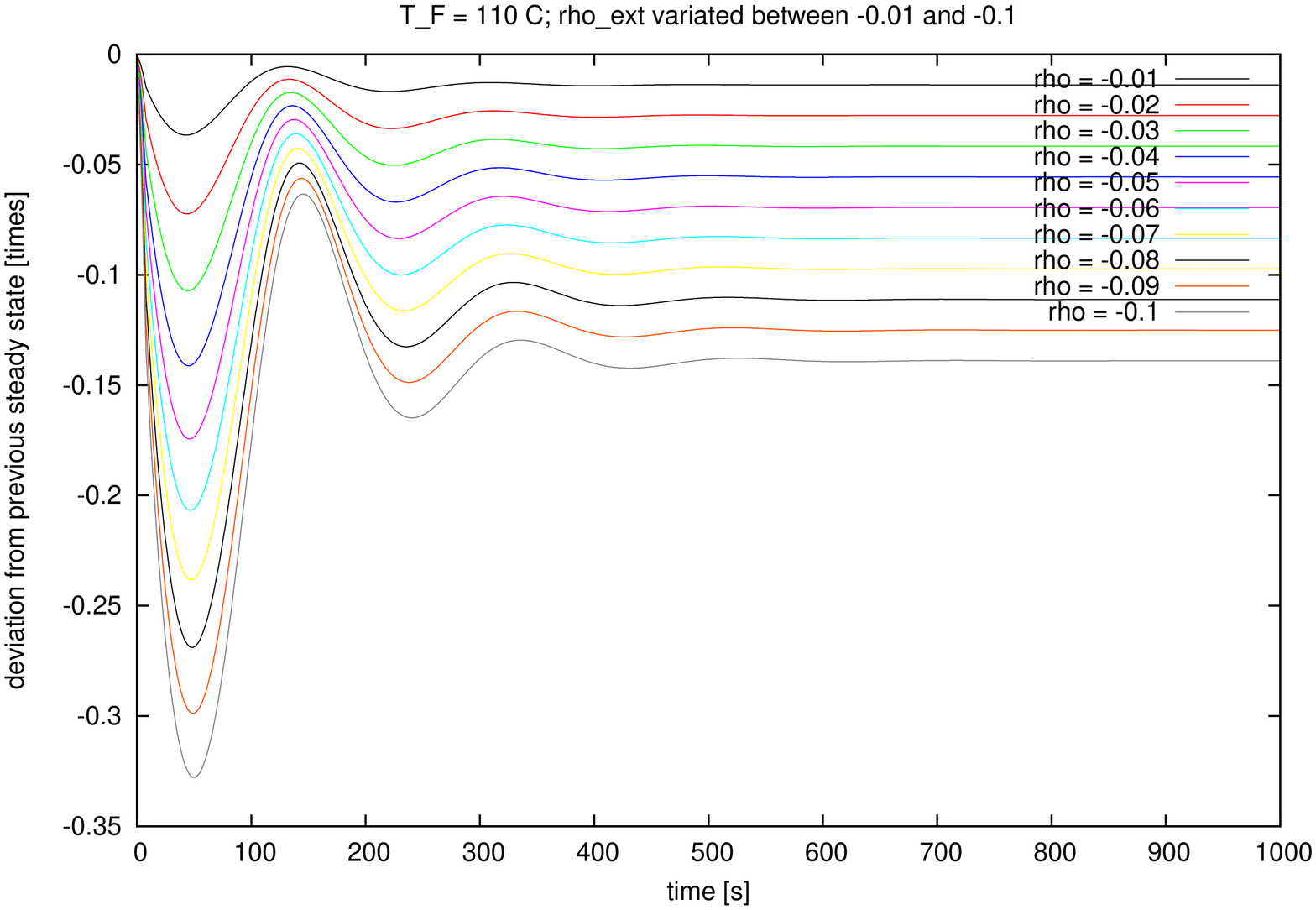}
\caption{Fuel rod temperature: 110 $^{\circ}$C, negative reactivity changes}
\label{default}
\end{center}
\end{figure}
\begin{figure}[htbp]
\begin{center}
\includegraphics[width=3.5in,angle=-90]{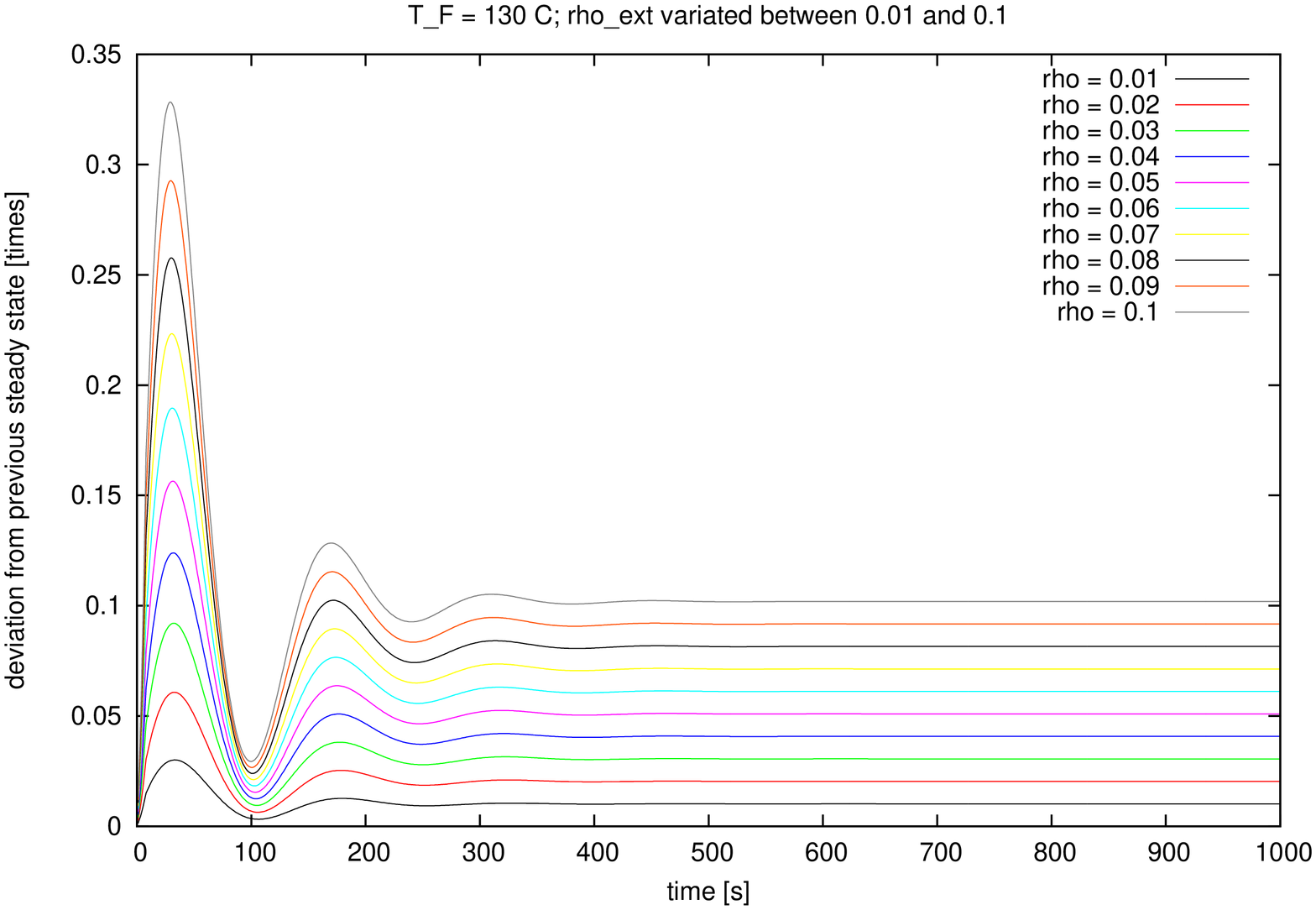}
\caption{Fuel rod temperature: 130 $^{\circ}$C, positive reactivity changes}
\label{default}
\end{center}
\end{figure}
\begin{figure}[htbp]
\begin{center}
\includegraphics[width=3.5in,angle=-90]{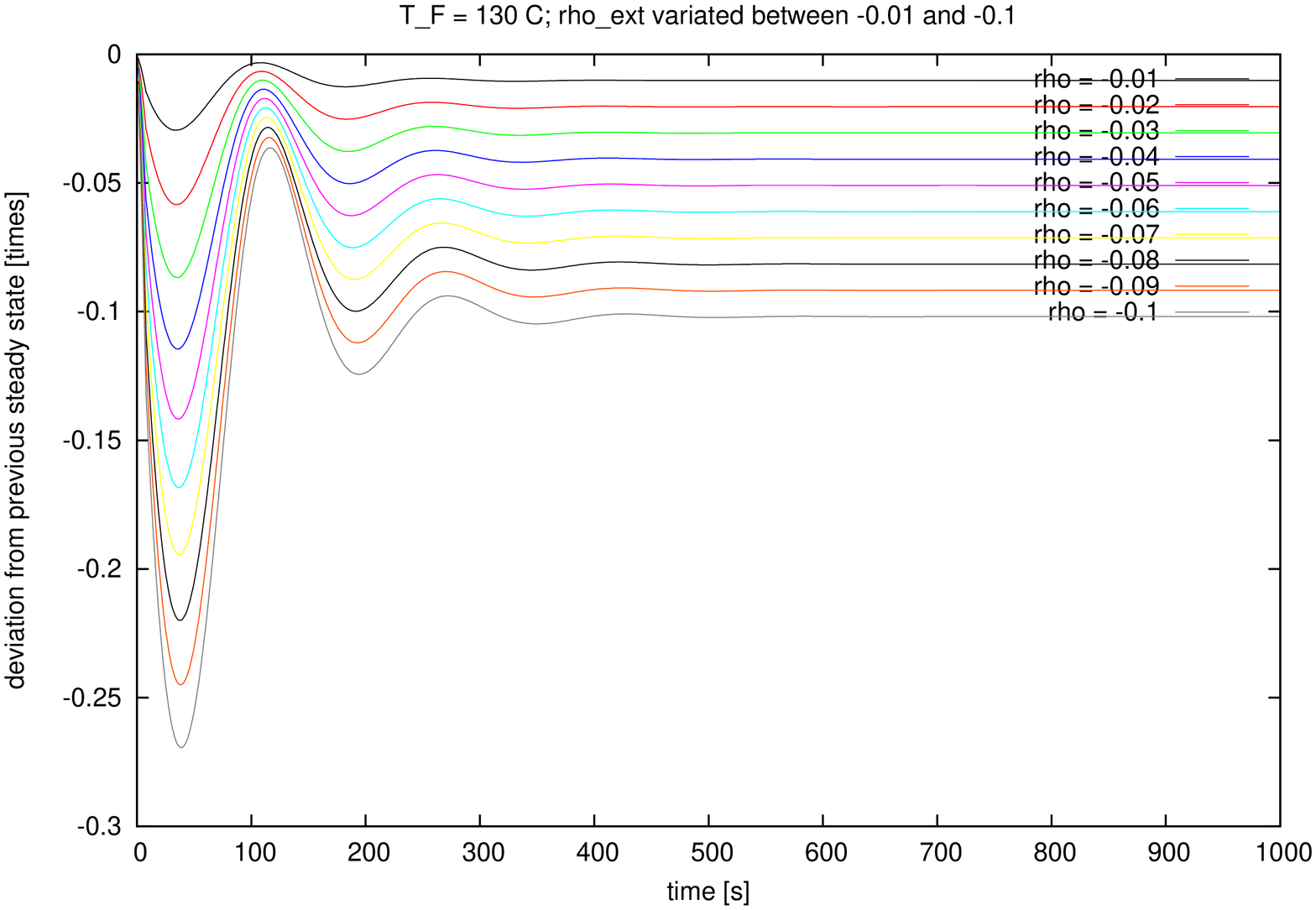}
\caption{Fuel rod temperature: 130 $^{\circ}$C, negative reactivity changes}
\label{default}
\end{center}
\end{figure}
\begin{figure}[htbp]
\begin{center}
\includegraphics[width=3.5in,angle=-90]{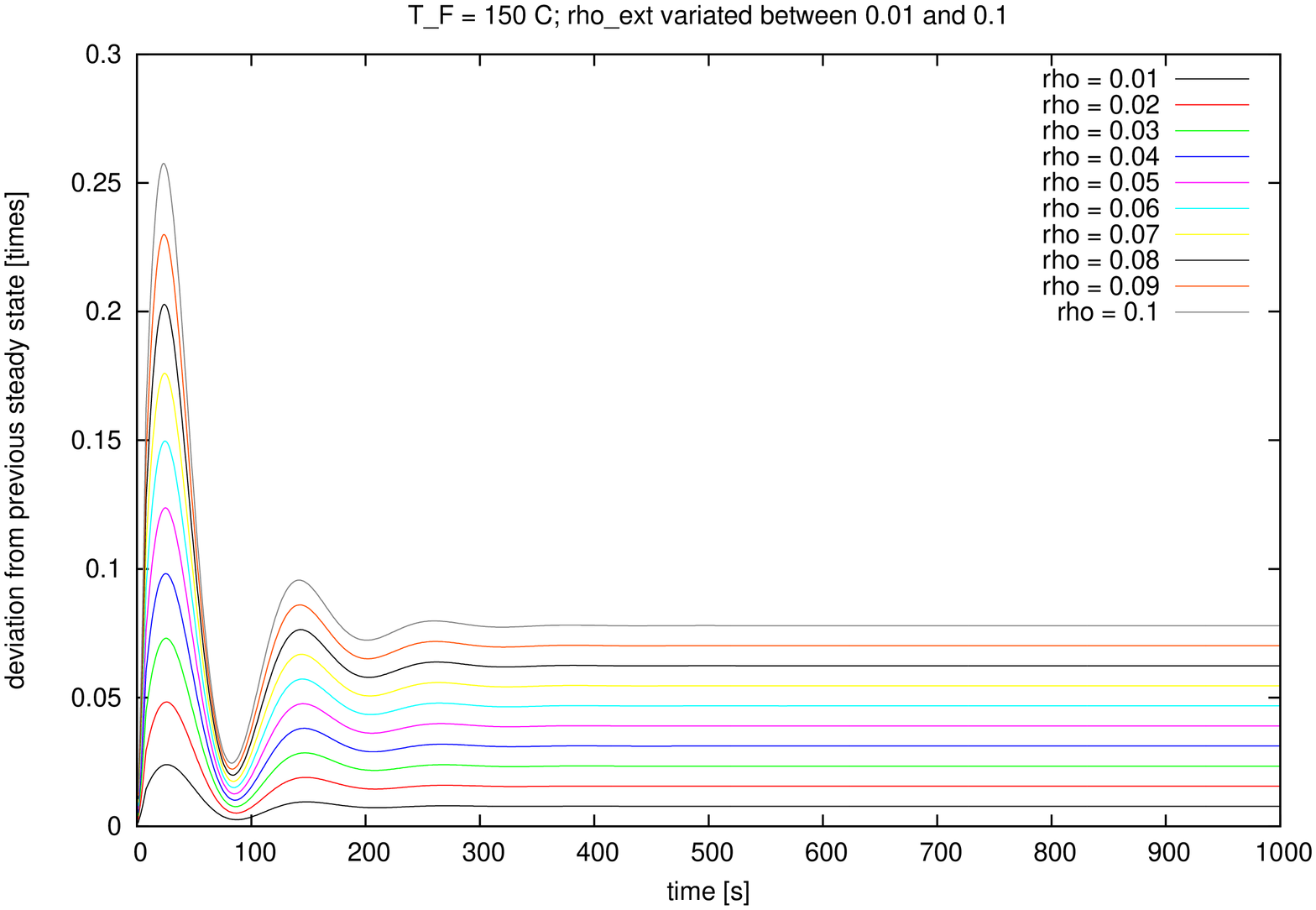}
\caption{Fuel rod temperature: 150 $^{\circ}$C, positive reactivity changes}
\label{default}
\end{center}
\end{figure}
\begin{figure}[htbp]
\begin{center}
\includegraphics[width=3.5in,angle=-90]{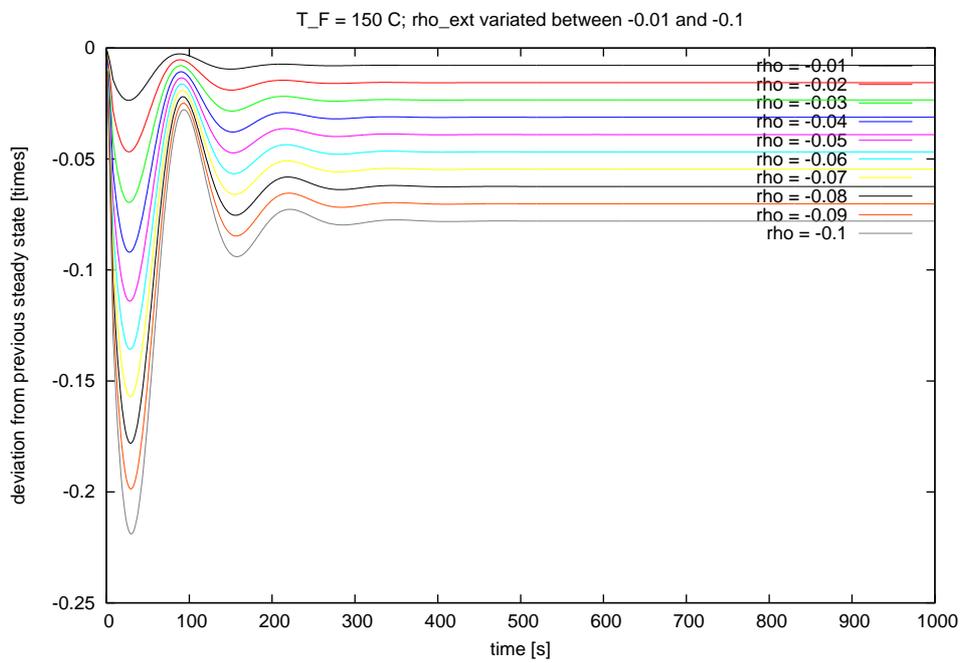}
\caption{Fuel rod temperature: 150 $^{\circ}$C, negative reactivity changes}
\label{default}
\end{center}
\end{figure}

\newpage

\begin{thebibliography}{}

\bibitem{triganss1}     Hugo M Dallea, Hugo M.; Pereirab, C; Souzaa, R. G. P; \emph{Neutronic calculation to the TRIGA Ipr-R1 reactor using the WIMSD4 and CITATION codes}, Annals of Nuclear Energy 29(8),2002
\bibitem{triganss2} Snoj, L.; Ravnik, M: \emph{Calculation of power density with MCNP in TRIGA reactor}, Nuclear Energy for New Europe 2006, Portoro?, Slovenia

\bibitem{triganss3} Snoj, Luka, et al. \emph{Calculation of kinetic parameters for mixed TRIGA cores with Monte Carlo}, Annals of Nuclear Energy 37.2 (2010): 223-229.

\bibitem{triganss4} Matsumoto, Tetsuo, and Nobuhiro Hayakawa, \emph{Benchmark analysis of TRIGA Mark II reactivity experiment using a continuous energy Monte Carlo code MCNP},  Journal of NUCLEAR SCIENCE and Technology 37.12 (2000): 1082-1087.

\bibitem{trigass1} Feltus, M. A; Miller, W. S.: \emph{Three-dimensional coupled kinetics/thermal- hydraulic benchmark TRIGA experiments}, Annals of Nuclear Energy 27:9 (2009)

\bibitem{trigass2} Reis, Patr'cia AL, et al.: \emph{Assessment of a RELAP5 model for the IPR-R1 TRIGA research reactor}, Annals of Nuclear Energy 37.10 (2010): 1341-1350.

\bibitem{dissrustam} Khan, R., et al.: \emph{Neutronics analysis of the current core of the TRIGA Mark II reactor Vienna.}, International conf. on Research Reactor Fuel Management RRFM 2010, Rabat, Morocco

\bibitem{xenon} Riede, J.; Boeck, H.: \emph{Spectrometrical in-core measurements of time-dependent Xenon-135 inventory in the TRIGA Mark II reactor Vienna}, publication in review

\bibitem{emendoerfer} Emendoerfer, D.; Hoecker, K.: \emph{Theorie der Kernreaktoren} Part I \& II, B.I. Wissenschaftsverlag Mannheim 1982
\bibitem{villa} Boeck, H.; Villa, M.: \emph{TRIGA Reactor Characteristics} AIAU 24309, June 2004
\bibitem{trigapdf} Boeck, H.; Villa, M.: \emph{The TRIGA Mark II reactor at Vienna University of Technology}, ATI Report, 2005
\bibitem{stacey} Stacey, W.; \emph{Nuclear Reactor Physics}, Wiley-VCH 2011

\end{thebibliography}
\end{document}